%% file: main_arXiv.tex
\documentclass[10pt,twocolumn,letterpaper]{article}

\usepackage[pagenumbers]{cvpr} 

\input{preamble}
\definecolor{cvprblue}{rgb}{0.21,0.49,0.74}
\usepackage[pagebackref,breaklinks,colorlinks,allcolors=cvprblue]{hyperref}

\usepackage{float}
\usepackage{caption}

\def\paperID{26} 
\def\confName{CVPR}
\def\confYear{2026}

\title{PhaseFlow4D: Physically Constrained 4D Beam Reconstruction via Feedback-Guided Latent Diffusion}

\author{Alexander Scheinker\\
Los Alamos National Laboratory\\
Los Alamos 87545, NM, USA\\
{\tt\small ascheink@lanl.gov}
\and
Alexander Plastun \quad Peter Ostroumov\\
Facility for Rare Isotopes Beams\\
Michigan State University, East Lansing, MI 48824, USA\\
{\tt\small \{plastun, ostroumo\}@frib.msu.edu}
}

\begin{document}

\twocolumn[{
    \maketitle
    \vspace{-2em}
    \centering
    \includegraphics[width=1.0\textwidth]{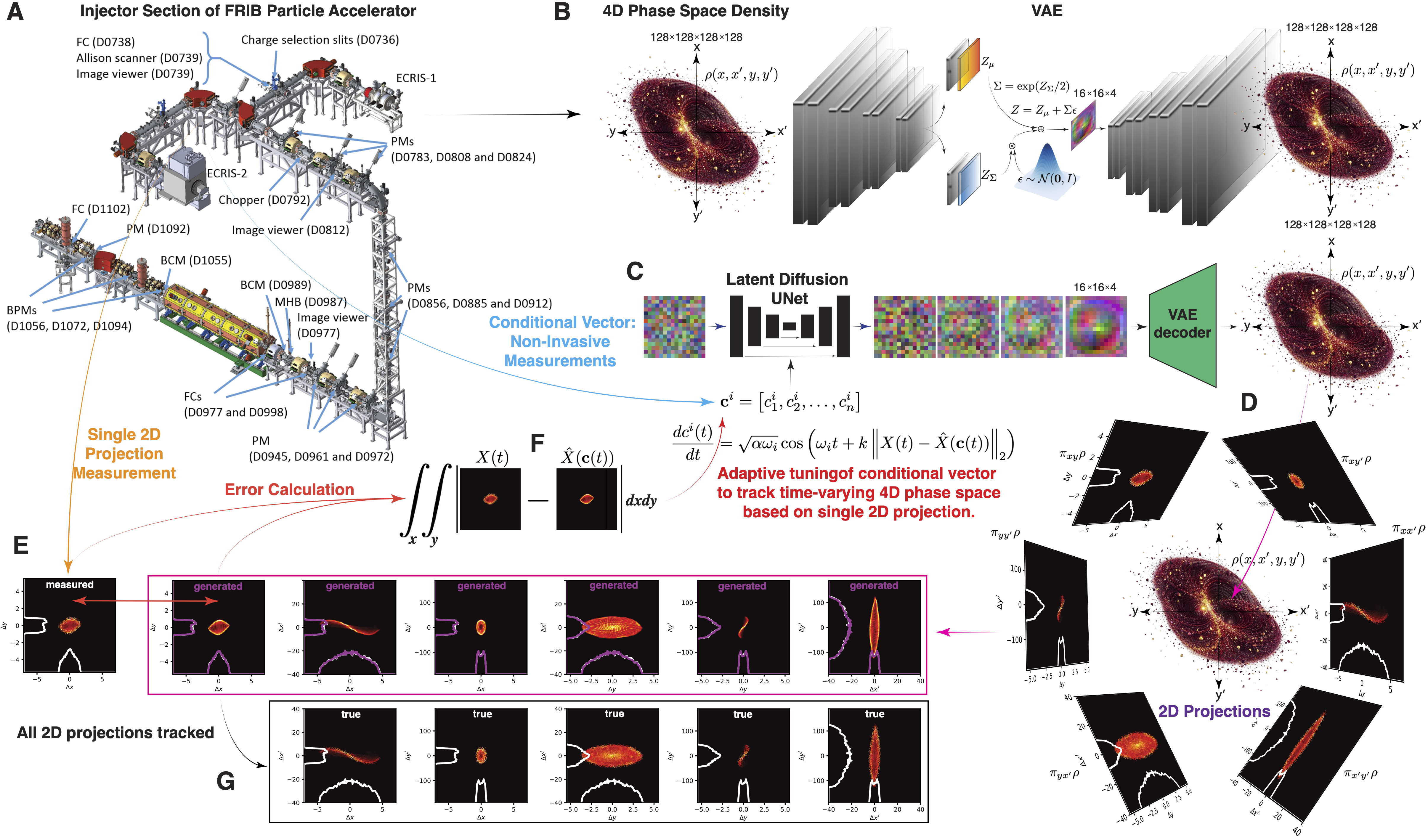}
    \captionof{figure}{{\bf A}: FRIB accelerator injector beam plasma source and charge selection system. {\bf B}: 4D phase space density $\rho(x,x',y,y')$ of beam initial conditions. Simulating complex space charge-dominated beam dynamics of 13 beam species computationally expensive (6 hours). 4D VAE encodes $128^4$ 4D density into a low-dimensional latent representation $16\times 16 \times 4$. {\bf C}: Latent diffusion conditional input based on beamline settings and non-invasive measurements. {\bf D}: Latent diffusion maps beamline conditions to full 4D phase space density from which physically consistent 2D projections are made. {\bf E}: A single generated 2D projection compared with measurement and the difference ({\bf F}) is minimized by adaptive conditional vector tuning. {\bf G}: Time-Varying 4D phase space distribution tracked based on 2D measurements.}
    \label{fig:overview}
    \vspace{0.5em}
}]

\input{sec/0_abstract} 
\input{sec/1_intro}
\input{sec/2_background}
\input{sec/3_method}
\input{sec/4_experiments}

\input{sec/5_conclusion}
{
    \small
    \bibliographystyle{IEEEtran}
    \bibliography{main}
}

\input{sec/ES_suppl}

\end{document}

%% file: sec/0_abstract.tex
\begin{abstract}
We address the problem of recovering a time-varying 4D distribution from a sparse sequence of 2D projections — analogous to novel-view synthesis from sparse cameras, but applied to the 4D transverse phase space density $\rho(x,p_x,y,p_y)$ of charged particle beams. Direct single shot measurement of this high-dimensional distribution is physically impossible in real particle accelerator systems; only limited 1D or 2D projections are accessible. We propose PhaseFlow4D, a feedback-guided latent diffusion model that reconstructs and tracks the full 4D phase space from incomplete 2D observations alone, with built-in hard physics constraints. Our core technical contribution is a 4D VAE whose decoder generates the full 4D phase space tensor, from which 2D projections are analytically computed and compared against 2D beam measurements. This projection-consistency constraint guarantees physical correctness by construction — not as a soft penalty, but as an architectural prior. An adaptive feedback loop then continuously tunes the conditioning vector of the latent diffusion model to track time-varying distributions online without retraining. We validate on multi-particle simulations of heavy-ion beams at the Facility for Rare Isotope Beams (FRIB), where full physics simulations require $\sim$6 hours on a 100-core HPC system. PhaseFlow4D achieves accurate 4D reconstructions $11000\times$ faster while faithfully tracking distribution shifts under time-varying source conditions — demonstrating that principled generative reconstruction under incomplete observations transfers robustly beyond visual domains.
\end{abstract}

%% file: sec/1_intro.tex
\section{Introduction}
\label{sec:intro}

Recovering a complete high-dimensional distribution from a sparse set of lower-dimensional projections is a fundamental inverse problem with applications spanning computational imaging, scientific simulation, and geometric reconstruction. In computer vision, this challenge underlies the problem of novel-view synthesis (NVS) and 3D scene recovery from sparse camera observations, the task of generating images of a scene from previously unobserved viewpoints. Early work explored image-based approaches such as view interpolation from pairs of images \cite{chen1993view}, as well as dense ray-based representations including light fields \cite{levoy1996light} and lumigraphs \cite{gortler1996lumigraph}. Recently, the NVS domain that has seen rapid progress through the combination of learned neural representations with generative modeling priors \cite{nerf2021,sparf2023,kerbl20233d,flowr2025,difix2025}. The central question common to all such settings is how to enforce consistency between the recovered representation and the available observations, while producing reconstructions that are physically or geometrically faithful rather than merely visually plausible. We address an instance of this problem that is strictly harder than its visual counterpart: reconstructing the full 4D transverse phase space density $\rho(x,p_x,y,p_y)$ of an intense charged particle beam — a time-varying, high-dimensional distribution observable only through a handful of 2D projection measurements.

\begin{figure}[h]
  \centering
  \includegraphics[width=1.0\linewidth]{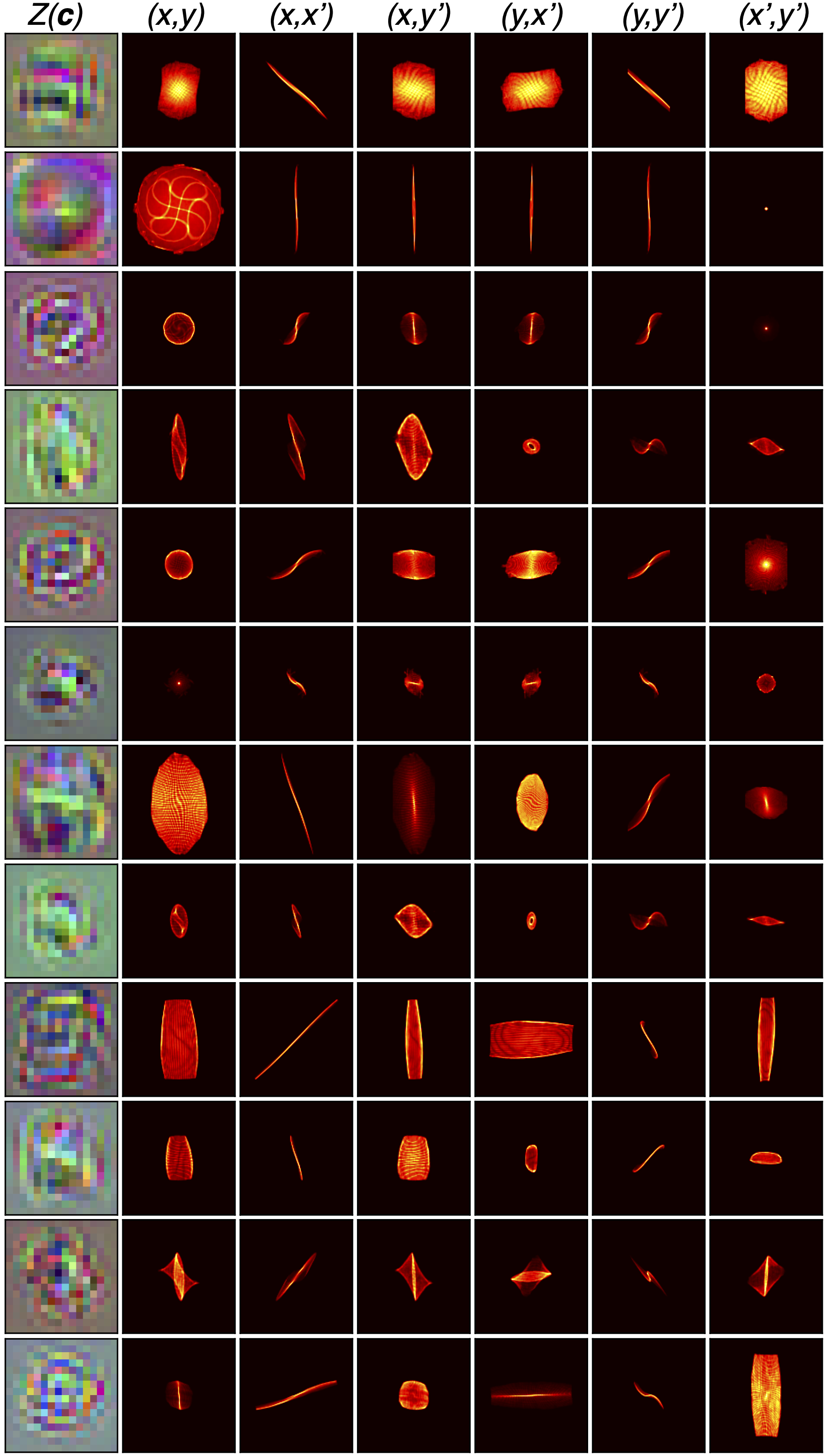}
  \caption{Examples of conditional latent diffusion-generated latent embeddings and all 6 unique 2D projections associated with the 4D phase space densities that those latent images are decoded to by the VAE's decoder.}
  \label{fig:GenExamples}
\end{figure}

Generative diffusion models have emerged as a powerful paradigm for high-dimensional reconstruction under incomplete observations. Foundational latent diffusion architectures \cite{rombach2022ldm} demonstrate that learning a compressed latent space via a variational autoencoder (VAE) enables stable, high-fidelity generation; conditional guidance mechanisms \cite{ho2022cfg,scheinker2024cdvae} further allow the denoising process to be steered toward specific targets at inference time. Building on these foundations, recent work has extended diffusion-based generation directly to 3D and 4D domains. DiFix3D+ \cite{difix2025} demonstrates that a single-step diffusion model can substantially improve artifact-corrupted 3D reconstructions by acting as a learned prior over plausible geometry, yielding metric-accurate improvements without iterative optimization. Similarly, Gen3C \cite{gen3c2025} shows that 3D-informed conditioning enables temporally consistent video generation with precise camera control, revealing how structural 3D priors propagate faithfully across time. Object-X \cite{objectx2025} further extends the generative paradigm to multi-modal 3D object representations, underscoring the generality of learned latent spaces as reconstruction intermediaries across diverse observation modalities.

A parallel and complementary line of work addresses the challenge of reconstruction under sparse or degraded observations. SPARF \cite{sparf2023} establishes that neural radiance field optimization from as few as three input images — with noisy camera poses — is feasible when multi-view geometric consistency constraints are folded into the training objective. This principle, that learned representations should be internally consistent with respect to projections under known geometric transformations, is central to our own design. FlowR \cite{flowr2025} takes this further, framing the problem of going from sparse to dense 3D reconstructions as a generative flow process that progressively refines incomplete point sets into geometrically complete structures. At the diffusion-for-3D-geometry intersection, P2P-Bridge \cite{p2pbridge2024} reformulates point cloud denoising as a Schr\"{o}dinger bridge problem, learning an optimal transport plan from noisy to clean 3D point sets — a formulation that shares our intuition that geometry recovery should be cast as a principled probabilistic mapping rather than a regression problem.

Despite this progress, existing generative reconstruction methods share a critical implicit assumption: observations consist of rendered images or depth maps from calibrated cameras. In scientific and engineering domains, however, the available measurements are often projections of the underlying distribution in a strictly physical sense — integrals of the full-dimensional density along unmeasured coordinates. No camera or view-synthesis prior applies. For charged particle beams in accelerator systems such as FRIB, direct 4D measurement of the transverse phase space $\rho(x,p_x,y,p_y)$ is physically impossible: the system is destructive, high-dimensional, and operates on microsecond timescales. Only 2D marginal projections (beam profile images) are observable, and the beam distribution itself evolves continuously under time-varying source conditions. This setting calls for a reconstruction paradigm centered on metric accuracy, projection consistency, and adaptive online tracking — precisely the goals motivating this workshop — but currently unaddressed by existing methods. 

The most closely related prior work for charged particle beams \cite{scheinker2025physics} used a VAE to map a 2D beam projection to a low-resolution 6D phase space tensor $\rho(x,p_x,y,p_y,z,p_z)$, followed by a super-resolution diffusion model to refine individual 2D projections from $32\times 32$ to $256\times 256$. This two-stage design is conceptually similar to FlowR, in that a generative model is used to upscale an initially coarse reconstruction. Its key limitation, however, is that the super-resolution diffusion model operates on individual 2D projections independently, breaking the hard physics constraints enforced by the VAE: the refined projections are no longer guaranteed to be marginals of any single consistent 4D or 6D distribution.

We propose \textbf{PhaseFlow4D}, a feedback-guided latent diffusion framework that avoids this inconsistency by maintaining a single physically consistent representation throughout. We focus on the 4D transverse phase space $(x,p_x,y,p_y)$ for two reasons. First, GPU memory constraints make single-pass generation of full-resolution tensors prohibitive in 6D: generating $128^4$ tensors on an H100 is already at the limit of feasibility, and scaling to $128^6$ would require dedicated HPC resources beyond the scope of this study. Second, the 4D transverse plane captures the dominant dynamical degrees of freedom in the accelerator systems we consider: the longitudinal coordinates $(z,p_z)$ are tightly controlled by the source voltage and are effectively known, so recovering $(x,p_x,y,p_y)$ is sufficient to reconstruct the full 6D distribution in practice.

Our approach introduces a 4D VAE whose decoder generates the complete phase space tensor, from which 2D projections are computed analytically and can then be compared against real beam measurements. A high level view of our approach is shown in Figure \ref{fig:overview}. This projection-consistency constraint acts as a hard architectural prior rather than a soft penalty, guaranteeing by construction that all marginal projections of the reconstructed distribution are physically correct. An adaptive feedback mechanism then continuously tunes the conditioning vector of the latent diffusion model — analogous in spirit to the pose-NeRF joint refinement of SPARF \cite{sparf2023} and the iterative artifact correction of DiFix3D+ \cite{difix2025} — enabling online tracking of time-varying distributions without retraining. Validated on multi-particle heavy-ion simulations at FRIB, PhaseFlow4D achieves accurate 4D reconstructions $11000×\times$ faster than the physics-based simulator, while faithfully tracking distribution shifts driven by time-varying ECR source conditions. Our results demonstrate that the core principles of generative reconstruction under incomplete observations — learned latent spaces, projection consistency, and adaptive conditioning — transfer robustly beyond the visual domain, opening a new frontier for AI-driven scientific diagnostics. Some examples of the latent images and the 2D projections of the resulting 4D phase space density distribution are shown in Figure \ref{fig:GenExamples}.

Our contributions can be summarized as follows.
\begin{itemize}
    \item We have developed a real-time 4D phase space diagnostic for charged particle beams requiring only single-shot 2D measurements.
    \item Our generative model has built in hard physics constraints in the form of projection-consistency.
    \item Our adaptive conditional guidance approach enables our diffusion model to track time-varying distributions.
\end{itemize}

%% file: sec/2_background.tex
\section{Background and Related Work}
\label{sec:related}

\paragraph{Latent diffusion models.}
Latent diffusion models (LDMs) \cite{rombach2022ldm} factorize generative modeling into two stages: a variational autoencoder (VAE) that compresses high-dimensional data into a compact latent space, and a denoising diffusion model that learns the prior over that latent space. Conditioning mechanisms, most notably classifier-free guidance (CFG) \cite{ho2022cfg}, allow the generative process to be steered at inference time toward a desired target by interpolating between conditional and unconditional score estimates.  More recent work has explored \emph{inference-time} optimization of the conditioning signal itself: rather than fixing the condition at the start of sampling, the condition (or the initial noise) is refined online using verifier feedback \cite{ma2025inference}, enabling substantial quality improvements without retraining the model.  PhaseFlow4D adopts a closely related philosophy: we treat the
conditioning vector $\mathbf{c}$ as a free variable to be optimized at deployment time, driven not by a perceptual verifier but by a physically grounded projection-consistency error.

\paragraph{Generative models for 3D/4D reconstruction.}
Diffusion-based priors have proven effective for 3D reconstruction under incomplete
observations.  DiFix3D+ \cite{difix2025} shows that a single-step diffusion model can
restore artifact-corrupted radiance fields by acting as a learned geometric prior, achieving metric-accurate improvements without iterative scene optimization.  Gen3C \cite{gen3c2025} demonstrates that structural 3D conditioning propagates faithfully across time, enabling world-consistent video generation with precise camera control.  Object-X \cite{objectx2025} extends learned latent representations to multi-modal 3D object recovery, underscoring the generality of VAE-based intermediaries across diverse observation modalities.  In the sparse-observation regime, SPARF \cite{sparf2023} establishes that joint pose--NeRF optimization from as few as three input images is feasible when multi-view projection-consistency constraints are embedded in the training objective — a principle directly mirrored in our projection-consistency VAE loss.  FlowR \cite{flowr2025} frames sparse-to-dense 3D reconstruction as a generative flow, progressively completing a partial point set; like FlowR, our approach uses a generative model to recover a high-fidelity representation from an initially incomplete observation, but our ``observations'' are physical marginal projections rather than camera images.  P2P-Bridge \cite{p2pbridge2024} reformulates point-cloud denoising as a Schr\"{o}dinger bridge between noisy and clean point sets, reinforcing the broader theme that geometry recovery is best cast as principled probabilistic mapping rather
than direct regression.

\paragraph{Phase space reconstruction for charged particle beams.}
A multi-modal conditional diffusion model was previously developed to track 2D projections of a charged particle beam's 6D phase space distribution \cite{scheinker2024cdvae}, but that approach conditionally generated single 2D projections thus lacking projection-consistency constraints. The most closely related prior work for particle accelerators \cite{scheinker2025physics} used a VAE to map a single 2D beam image to a low-resolution 6D phase space tensor $\rho(x,p_x,y,p_y,z,p_z)$ at $32^6$ voxels, followed by a super-resolution diffusion model to refine individual 2D projections from $32\times32$ to $256\times256$ pixels. The key limitation is that the super-resolution stage operates on individual projections independently, breaking the hard physics constraint imposed by the VAE: the refined images are no longer guaranteed to be marginals of any single consistent phase space distribution. PhaseFlow4D avoids this inconsistency entirely by maintaining a single physically consistent 4D representation throughout, generating the full tensor in one pass and computing all projections analytically from it. In what follows, as is typical in the accelerator community, instead of $p_x$, and $p_y$ we will refer to $x'=p_x/p_z$ and $y'=p_y/p_z$ which represents the divergence of the beam from a straight parallel path along the accelerator in the $z$-direction, where in our case $p_z$ is fixed for all particles in the beam which is a continuous stream uniform in $z$.
 
\paragraph{Model-free adaptive feedback and extremum seeking.}
Bounded extremum seeking (ES) \cite{ref_ES_1,scheinker2017es} is a class of model-free, gradient-free real-time stabilization and optimization methods that drive a dynamical system toward the extremum of a cost or Lyapunov function using only online measurements of that analytically unknown function.  ES requires no model of the system, no derivatives, and no retraining, making it well suited to tracking slowly time-varying optima in physical systems.  Here we use ES to continuously minimize the $\ell_2$ projection-consistency error between the observed 2D beam measurement and the 2D projection of our generated distribution, treating the conditioning vector $\mathbf{c}$ as the tunable parameter.  This is conceptually related to inference-time search over conditioning inputs \cite{ma2025inference}, but operates in a closed-loop, real-time setting where the ``verifier'' is a live physical measurement rather than a learned discriminator.

%% file: sec/3_method.tex
\section{Method}
\label{sec:method}

\begin{figure}[h]
  \centering
  \includegraphics[width=1.0\linewidth]{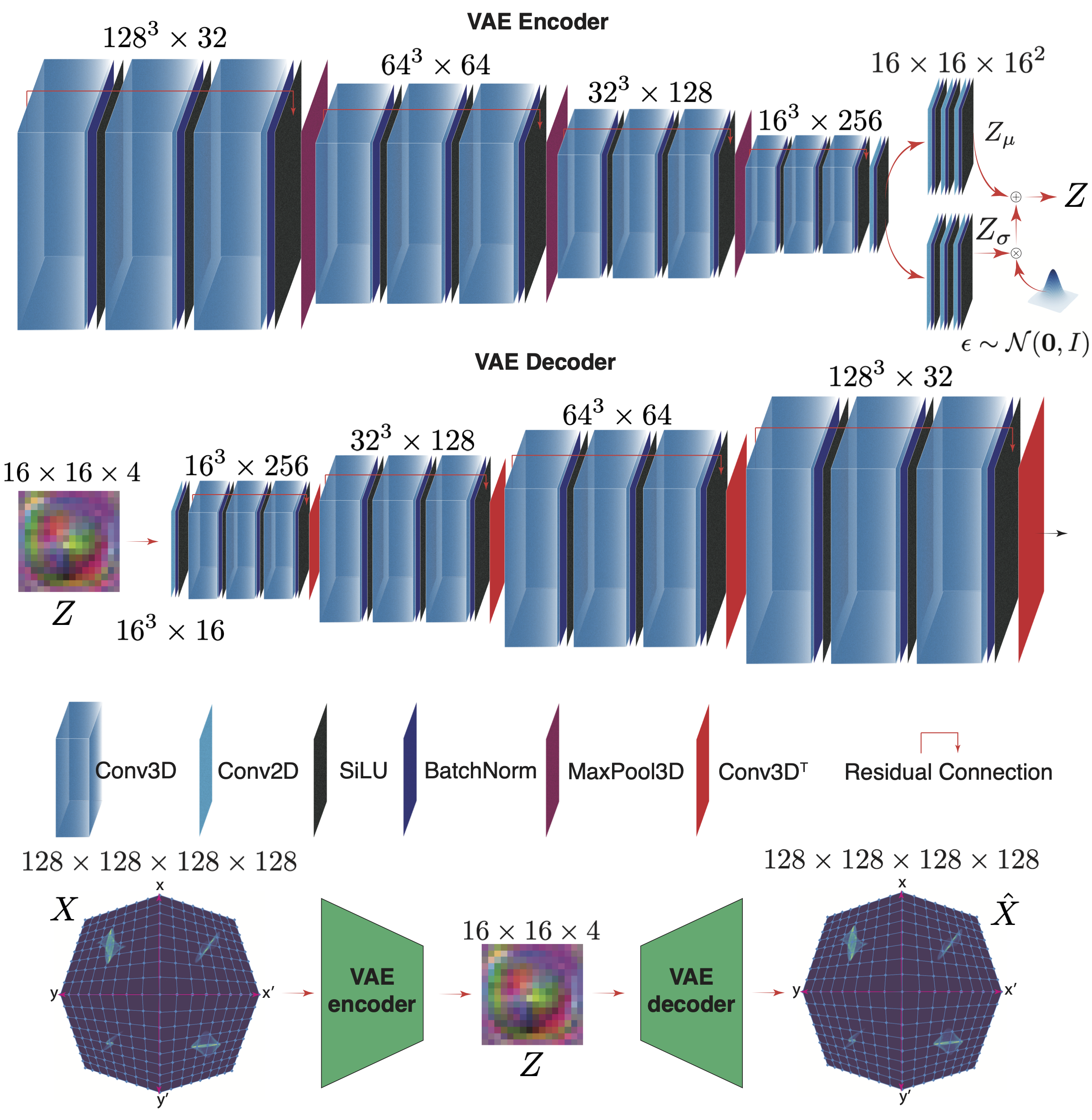}
  \caption{4D VAE architecture. Top: the encoder compresses the $128^4$ phase space tensor to a compact latent code $\mathbf{z}$. Bottom: the decoder reconstructs $\hat{X}$, from which all 2D marginal projections can be computed analytically and compared to the ground-truth projections during training. }
  \label{fig:vae}
\end{figure}

PhaseFlow4D combines three components: (1) a 4D VAE that encodes and decodes the full
transverse phase space density with a hard projection-consistency constraint, (2) a
conditional latent diffusion model that maps an accelerator condition vector to the
corresponding VAE latent code, and (3) an extremum-seeking feedback loop that
adaptively tunes the condition vector at deployment time to track a time-varying beam
distribution using only live 2D projection observations.  Figure~\ref{fig:overview}
provides a high-level schematic of the full system.

\subsection{4D Variational Autoencoder with Projection Consistency}
\label{sec:vae}

Let $X \in \mathbb{R}^{N \times N \times N \times N}$ denote a discretized 4D transverse
phase space density on a grid of resolution $N^4$ (we use $N=128$ throughout).
The VAE encoder $q_\phi(\mathbf{z} \mid X)$ maps $X$ to a latent code
$\mathbf{z} \in \mathbb{R}^{16\times 16 \times 4}$, and the decoder $p_\theta(X \mid \mathbf{z})$ maps
$\mathbf{z}$ back to a reconstructed density $\hat{X}$.

\begin{figure}[h]
  \centering
  \includegraphics[width=1.0\linewidth]{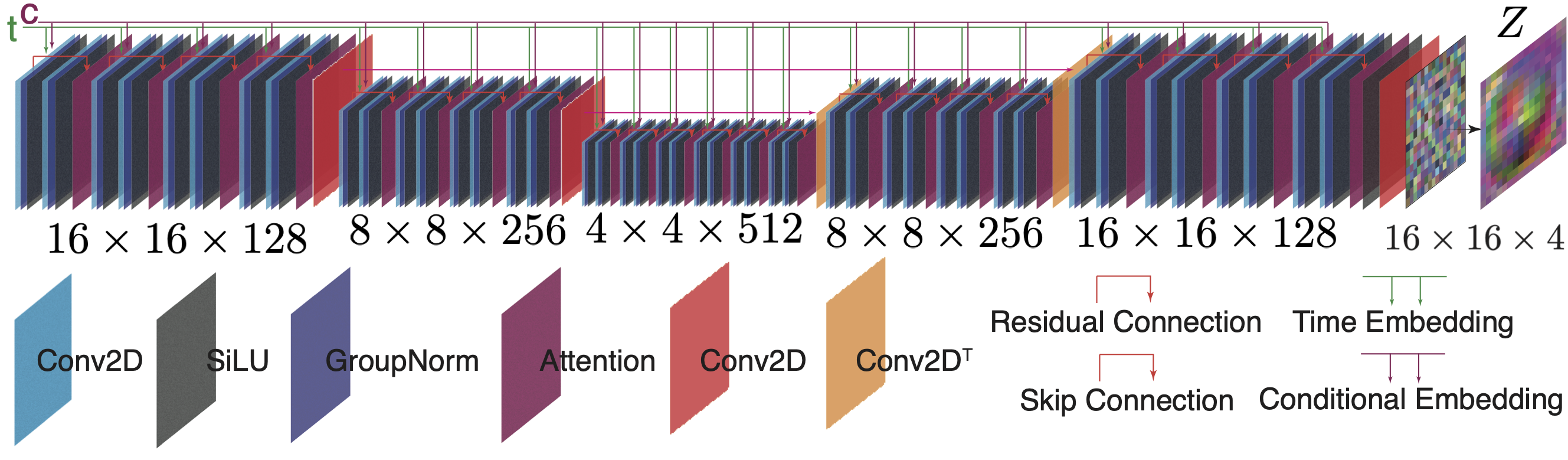}
  \caption{The conditional latent diffusion architecture is a standard U-Net approach with 3 residual blocks at each resolution, GroupNorm, and attention and 100 denoising steps.}
  \label{fig:cDiff}
\end{figure}

The key architectural constraint is that the decoder output $\hat{X}$ is used to
compute all six 2D marginal projections analytically by summing over the appropriate
pairs of dimensions:
\begin{equation}
  \pi_{ij}\hat{X} = \sum_{k,l \neq i,j} \hat{X}, \quad i,j \in \{ x, x', y, y' \}, \quad i\neq j.
\end{equation}
The training loss combines the standard VAE evidence lower bound (ELBO) with a
density prediction term:
\begin{equation}
  \mathcal{L}_{\mathrm{VAE}} = \mathcal{L}_{\mathrm{ELBO}}
  + \lambda \sum_{(i,j,k,l)} \bigl\| X - \hat{X} \bigr\|_2^2.
\end{equation}
This constraint is a \emph{hard architectural prior}: because in application all 2D projections will be computed from the same decoder output $\hat{X}$, physical consistency across projections is guaranteed by construction at every forward pass — not enforced as a soft penalty that can be violated at inference. Furthermore, by learning the actual 4D distribution, the model's predictions can be sampled and used as inputs to particle tracking codes to continue predicting the beam's evolution through further stages of the particle accelerator.

The VAE's encoder and decoder both utilize a 3 layer deep residual block at each resolution, as shown in Figure \ref{fig:vae}. At the input of the VAE, the $128^4$ tensor is interpreted as a $128^3$ 3D volume with $128$ channels and 3D convolutions are performed throughout, until a $16^4$ object is re-shaped into a $16\times 16$ image with $16^2$ channels before finally converting to a $16\times 16\times 4$ latent representation via 2D convolutions.

\subsection{Conditional Latent Diffusion Model}
\label{sec:ldm}

Given the trained VAE, we train a conditional denoising diffusion probabilistic model (DDPM) \cite{ho2020ddpm} in the latent space of the VAE.  The model takes as input a condition vector $\mathbf{c}=[c^1,\dots,c^{20}]\in\mathbb{R}^{20}$, where $c^1,\dots ,c^{13}$ is a one-hot encoding of 13 different beam species that are simultaneously being transported through the accelerator, $c^{14}$ encodes one of 23 locations along the accelerator,$c^{15}$ encodes a charge neutralization factor, which is a time-varying function of the source, and $c^{16},\dots,c^{20}$ encode the settings of 5 magnets along the beamline. The magnets are 2 solenoids and 3 quadrupoles which are adjustable and used to focus the beam. This conditional vector is used to guide the generative diffusion process to generate the latent code $\mathbf{z}(\mathbf{c})$ corresponding to the equilibrium 4D phase space distribution associated with those accelerator conditions. Conditioning is implemented via cross-attention in the denoising U-Net, following the standard LDM formulation \cite{rombach2022ldm}.

The full generative pipeline at inference is:
\begin{equation}
  \mathbf{c}
  \;\xrightarrow{\text{LDM}}\;
  \hat{\mathbf{z}}(\mathbf{c})
  \;\xrightarrow{\text{VAE decoder}}\;
  \hat{X}(\mathbf{c})
  \;\xrightarrow{\text{projection}}\;
  \pi_{ij}(\mathbf{c}).
\end{equation}
The entire chain is forward-only at deployment: no gradients are computed, and no fine-tuning is performed.  Generation of a full $128^4$ phase space estimate takes approximately $2$ seconds on a single H100 GPU, with the 100 diffusion steps taking approximately 1.5 seconds and the VAE's decoder 0.5 seconds, compared to $\sim$6 hours for the TRACK physics simulation.

\begin{figure}[h]
  \centering
  \includegraphics[width=1.0\linewidth]{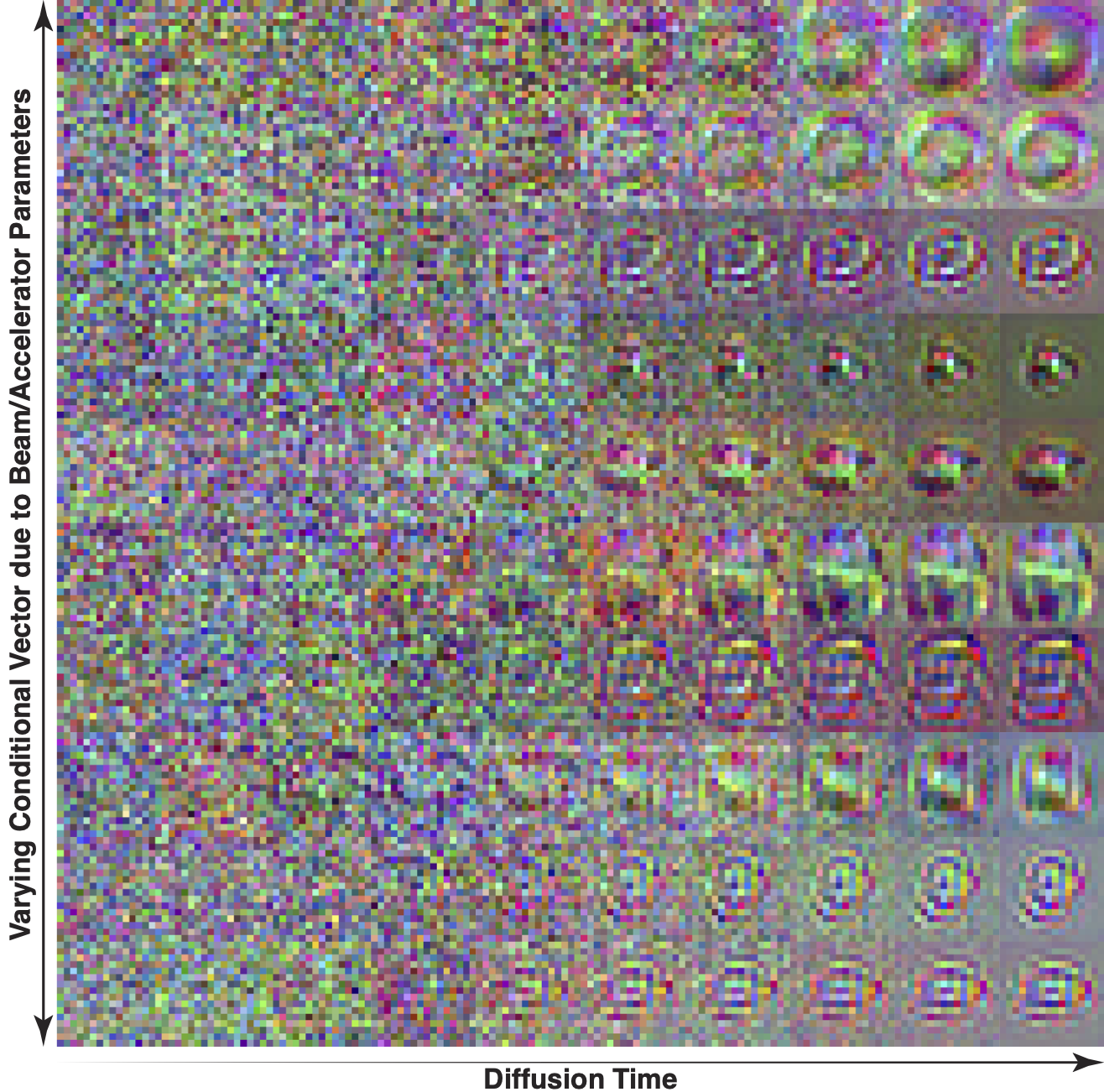}
  \caption{The conditional latent diffusion generative process is shown at various diffusion steps for different beam conditions. In this image, we show only the first three channels, as RGB values, of each $16 \times 16 \times 4$ latent images.}
  \label{fig:cDiffZGen}
\end{figure}

\subsection{Adaptive Condition Tuning via Extremum Seeking}
\label{sec:es}

In a real accelerator, the beam distribution $X(t)$ varies continuously as operating conditions drift — for example, due to charge neutralization fluctuations in the ECR plasma source.  We assume that at each time step $t$, we can observe only a single 2D projection of the beam, specifically the $(x,y)$ transverse profile image $\pi_{\mathrm{obs}}(t)$, as this is the one observable that is available in the FRIB front-end.

We formulate tracking as online minimization of the projection-consistency cost:
\begin{equation}
  J(\mathbf{c}(t),t) = \bigl\| \pi_{\mathrm{obs}}(t) - \pi_{xy}\hat{X}(\mathbf{c}(t)) \bigr\|_2^2.
\end{equation}
We minimize $J(\mathbf{c}(t),t)$ in real time using model-free extremum seeking (ES) \cite{scheinker2017es}.  ES requires no analytical model of the mapping $\mathbf{c} \mapsto J(\mathbf{c})$, no gradients, and no assumption that the landscape is fixed — it continuously tracks a time-varying optimum by injecting small dither perturbations and correlating the resulting cost fluctuations to infer a descent
direction.  The update rule takes the form:
\begin{equation}
    \frac{dc^i(t)}{dt} = \sqrt{\alpha\omega_i}\cos(\omega_i t + k J(\mathbf{c}(t),t)),
\end{equation}
where the dithering frequencies must be distinct with $\omega_i = \omega r_i$, where $r_i\neq r_j$ for all $i\neq j$. In the limit as $\omega \rightarrow \infty$, the average dynamics of this feedback loop are
\begin{equation}
    \frac{d\mathbf{c}(t)}{dt} = -\frac{k\alpha}{2}\nabla_\mathbf{c}J(\mathbf{c}(t),t),
\end{equation}
a gradient descent of the cost function. Recall that for any $T>0$, for the Hilbert space $L^2[0,T]$, a sequences of functions $\{f_i(t)\}$ is said to weakly converge to $f(t)$, denoted $f_i(t)\rightharpoonup f(t)$ if for any $g(t)\in L^2[0,T]$
\begin{equation}
    \lim_{i\rightarrow\infty} \langle f_i, g \rangle = \lim_{i\rightarrow\infty}\int_{0}^{T}f_igdt = \int_{0}^{T}fgdt = \langle f, g \rangle.
\end{equation}
The convergence proof can be found in \cite{ref_ES_1,scheinker2017es}, it essentially depends on the fact that for increasing $\omega_i$, $\cos(\omega_i t)\rightharpoonup 0$, $\cos(\omega_i t)\cos(\omega_j t)\rightharpoonup 0$ for $i\neq j$, and $\cos^2(\omega_i t)\rightharpoonup 1/2$.

\begin{figure}[h]
  \centering
  \includegraphics[width=1.0\linewidth]{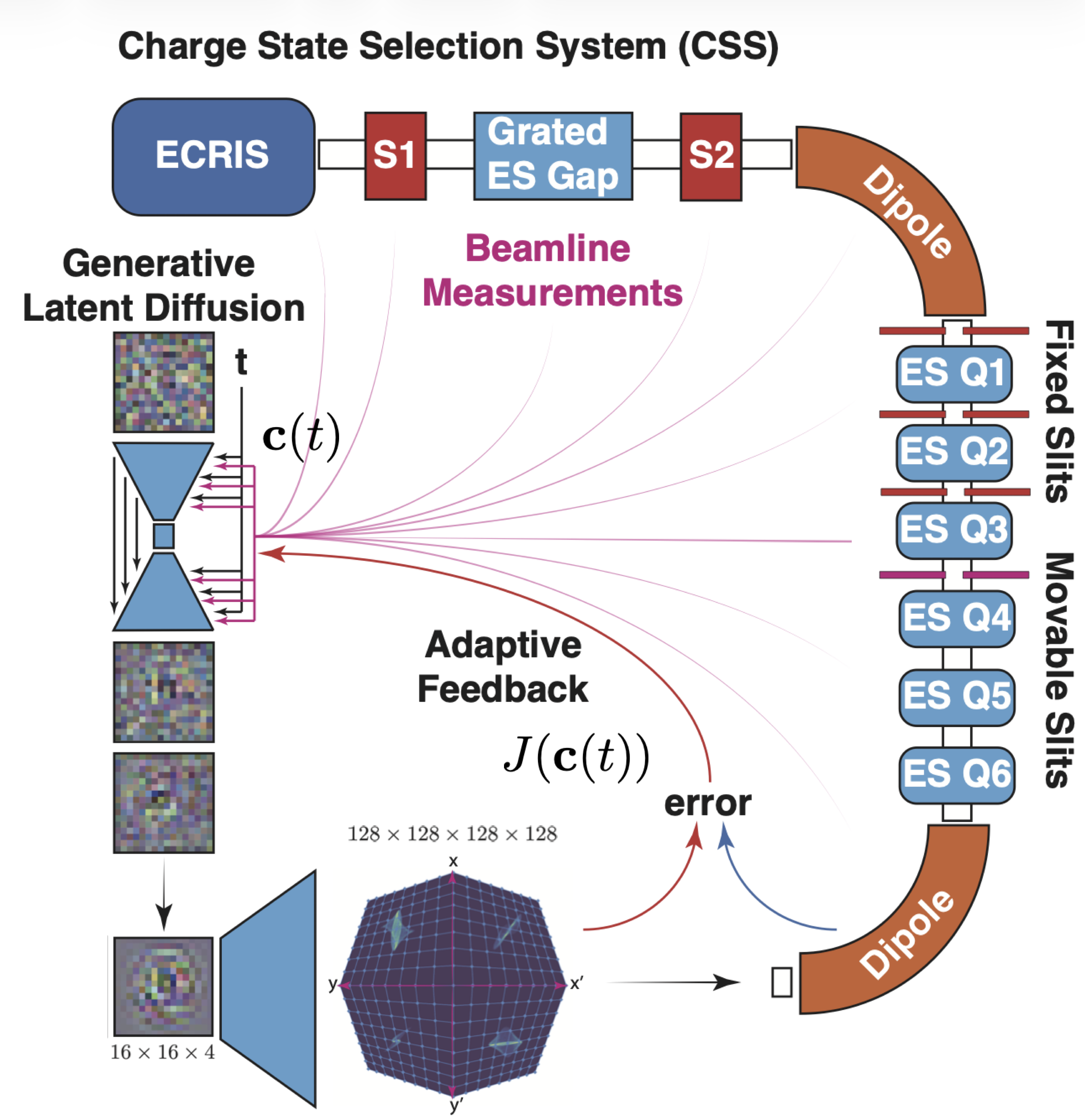}
  \caption{Overview of the the adaptive feedback setup.}
  \label{fig:CSS_ES}
\end{figure}

\begin{figure*}[h]
  \centering
  \includegraphics[width=1.0\linewidth]{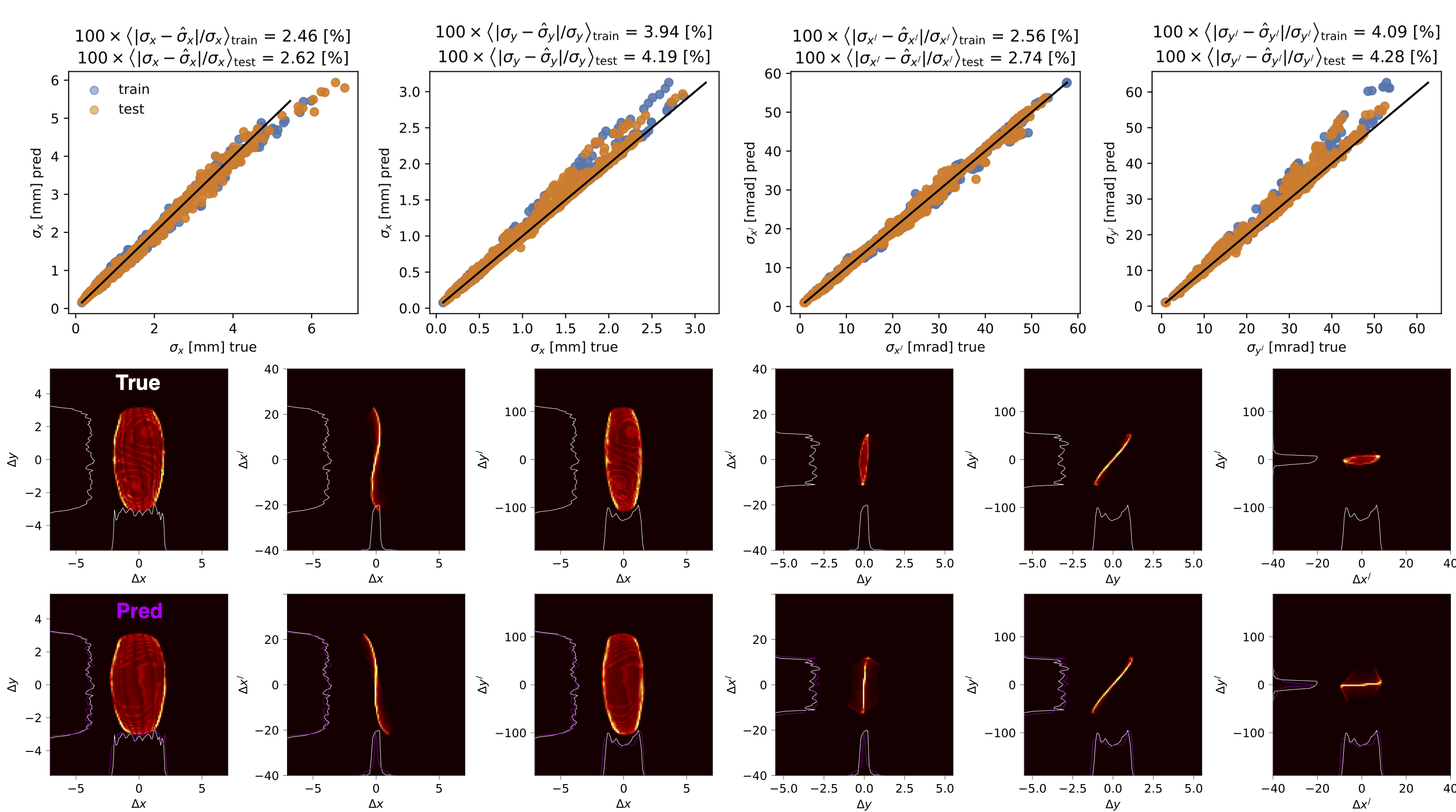}
  \caption{Top: Error statistics are shown for test and training data, where Gaussians have been fit to the 1D $(x,x',y,y')$ projections that were created from the generated 4D tensors. Bottom: One detailed example of all of the generated 2D projections relative to the true projections for a single test data point at the end of the beamline, where the test data input was only the conditional vector into the latent diffusion model.}
  \label{fig:train_test_sigmas}
\end{figure*}

The term $\alpha$ can be thought of as a dithering amplitude because at steady state, when $J$ is not changing, parameters undergo a steady state oscillation of the form
\begin{equation}
    c^i(t) \approx \sqrt{\frac{\alpha}{\omega_i}}\sin(\omega_i t),
\end{equation}
and $k$ can be thought of as a feedback gain. Although increasing either $\alpha$ or $k$ results in faster convergence, it is convenient to decrease $\alpha$, thereby limiting oscillation sizes, while increasing $k$ to maintain convergence rate.

This method which is implemented iteratively as a finite difference approximation according to
\begin{equation}
  c^i(n+1) = c^i(n) + \Delta_t \sqrt{\alpha\omega_i}\cos(\omega_i n \Delta_t + J(n)),
\end{equation}
where $J(n)$ represents $J(\mathbf{c}(n\Delta_t),n\Delta_t)$. Critically, the only information consumed by ES is the scalar cost $J$ evaluated at successive values of $\mathbf{c}$: it is entirely model-independent.

Because our 4D VAE enforces projection consistency by construction, minimizing the error on the \emph{single observable} projection $\pi_{xy}$ implicitly constrains all six 2D marginals of $\hat{X}$.  We therefore expect — and empirically verify — that accurately tracking the observable $(x,y)$ projection also causes PhaseFlow4D to track the remaining five unobserved projections, recovering the full 4D phase space distribution from a single 2D measurement stream. The adaptive setup is shown in Figure \ref{fig:CSS_ES}. As the ion source parameters drift, we only have access only to $\pi_{xy}(t)$, we compare it its projection that we generate from the 4D density, and continuously minimize $J(\mathbf{c}(t),t)$ in real time by tracking $\mathbf{c}(t)$.

%% file: sec/4_experiments.tex
\section{Experiments}
\label{sec:experiments}

\subsection{Dataset: FRIB Heavy-Ion Beam Simulations}
\label{sec:dataset}

All experiments use multi-particle simulations of 13 isotopes coming out of the FRIB ion source simultaneously, which creates a plasma beam composed of 7 Tin isotopes: $^{124}$Sn$^{22+}$-$^{124}$Sn$^{28+}$ and 6 Oxygen isotopes: $^{16}$O$^{+1}$-$^{16}$O$^{+6}$. It is important to simulate the transport of all the heavy-ions and the Oxygen through the charge selection section simultaneously by the TRACK code \cite{track_ref} to correctly account for 3D space charge effects. Only the $^{124}$Sn$^{26+}$ isotope is designed to travel along the center and survive through until the end as the beams come around a curve and are separated due to the different radius of curvature that each charge to mass ratio ion trajectory has. Each species is modeled by 300 thousand macroparticles and each simulation took approximately 6 hours on a 100-core machine. A total of 300,000 4D densities were used for model training. These densities were extracted from 1000 simulations with the full 4D phase space being recorded at 23 different locations along the beamline for each of the 13 charge states, resulting in $\sim$ 300 thousand 4D densities from which $\sim 1.8$ million 2D projections can be generated. Out of these, 270,000 4D densities (900 simulations) were used for model training, and 3,000 4D densities (10 held out simulations) were held out as validation data, and 27,000 4D densities (90 simulations) were used as test data.

\begin{figure*}[h]
  \centering
  \includegraphics[width=1.0\linewidth]{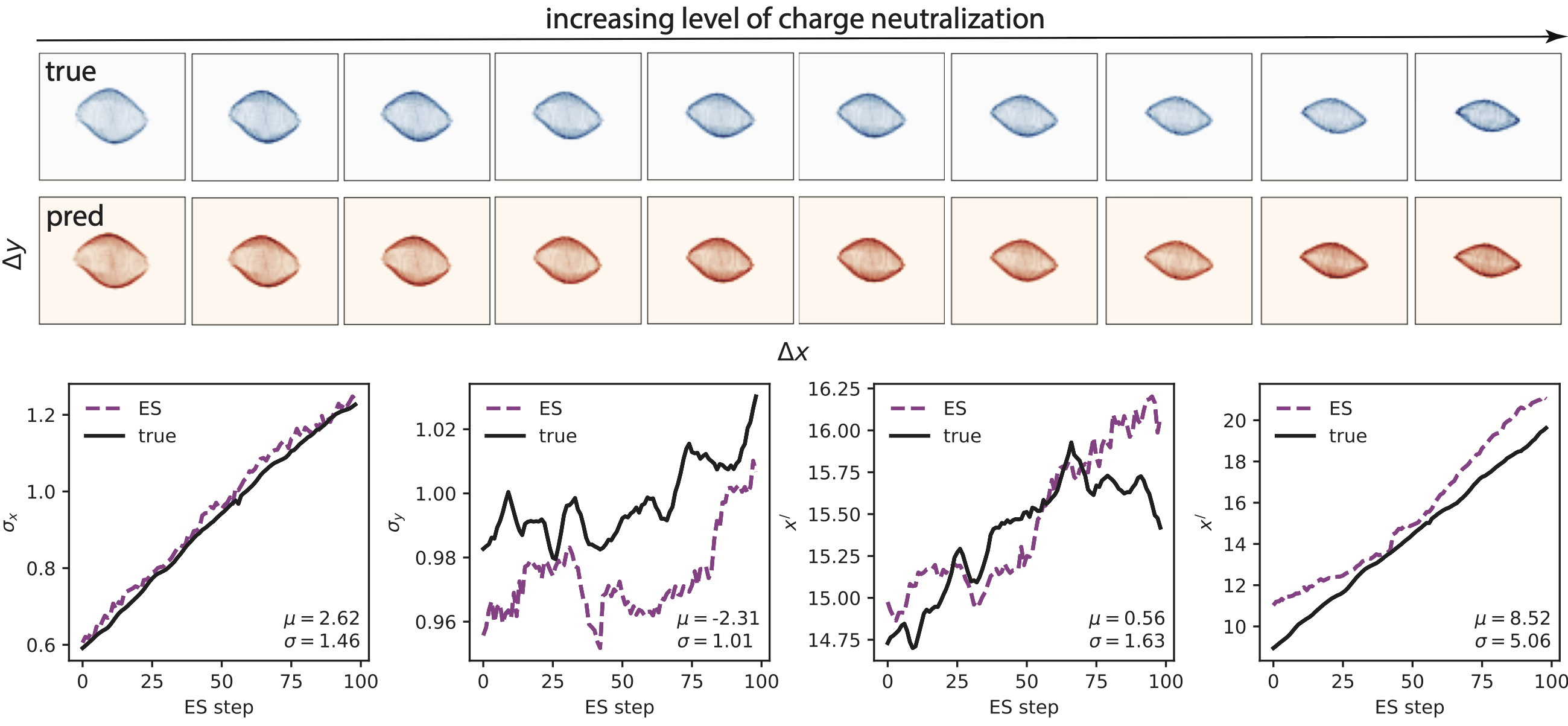}
  \caption{Top: The true $(x,y)$ projection of the beam is shown (blue) relative to the generated prediction (red) where tracking is performed by adaptive tuning of the diffusion conditional vector. Bottom: Although only the $(x,y)$ projection is available and used for tracking, the entire 4D phase space density is accurately tracked as seen by the true vs predicted $(\sigma_x,\sigma_{x'},\sigma_y, \sigma_{y'})$ fits of all 1D projections $(x,x',y,y')$ of the beam.}
  \label{fig:ES_Track_sigmas}
\end{figure*}

\begin{figure}[h]
  \centering
  \includegraphics[width=1.0\linewidth]{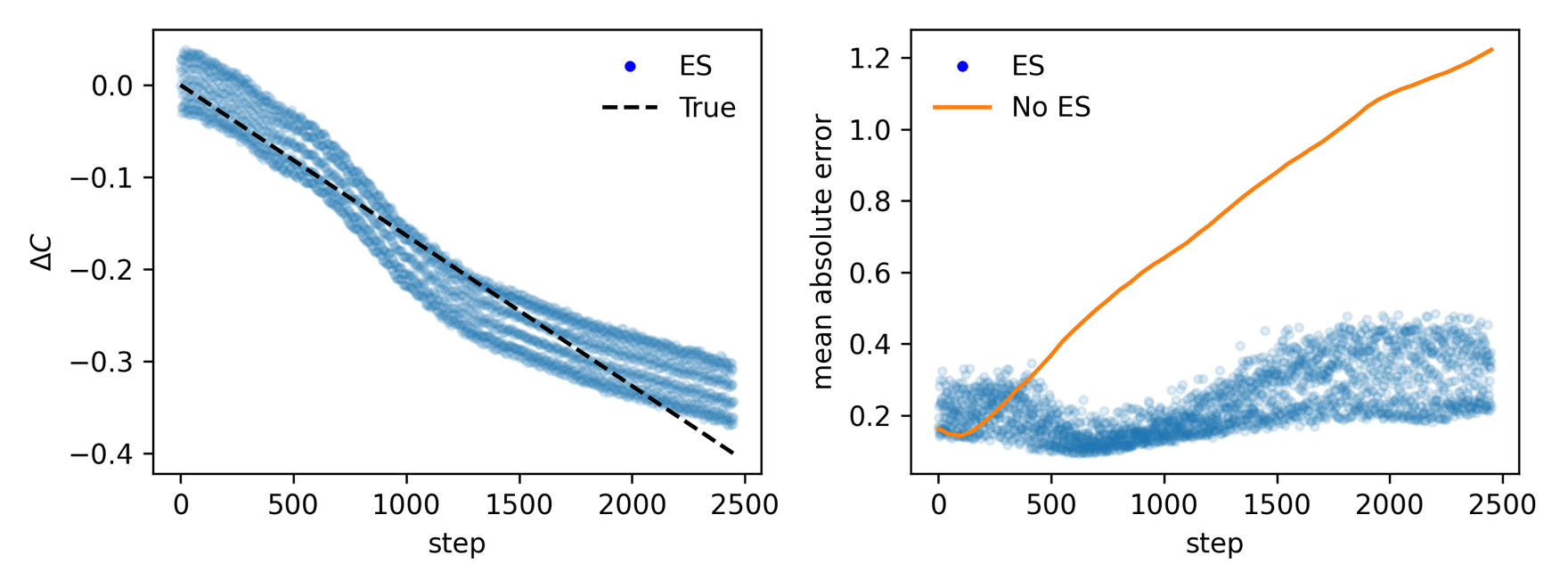}
  \caption{ES-based charge state tracking based on the $\pi_{xy}X(t)$ measurement (left) and error with and without tracking (right).}
  \label{fig:charge_state}
\end{figure}

In each simulation a uniformly distributed random perturbation was introduced to the two solenoid magnets and the 2 dipole magnets in the charge separation section whose schematic is shown in Figure \ref{fig:CSS_ES}, as well as to an overall charge neutralization factor which influences all of the beam species due to an imperfect vacuum and the presence of electrons in the plasma. After training, we re-generated all 4D distributions be passing their associated conditional vectors into the latent diffusion model, decoding the resulting latent image, and then creating various 2D and 1D projections to quantify the results in terms that are improtant to accelerator beam physicists. Figure \ref{fig:train_test_sigmas} shows the train vs test statistics where all 4D tensors were projected to their various 2D images and then further projected to 1D so that Gaussians could be fit to measure $\sigma_x$, $\sigma_{x'}$, $\sigma_y$, and $\sigma_{y'}$ of the beams relative to their true values, which are quantities of interest to beam physicists. A detailed test reconstruction is also shown relative to the true distribution.

For the adaptive tracking experiment, we constructed a time-varying sequence by parameterizing a smooth trajectory through condition space that models a realistic charge-neutralization drift scenario in the ECR source, starting from a known
calibrated condition and drifting. Throughout the drift we only have access the measurement $\pi_{xy}X(t)$. The results are shown in Figure \ref{fig:ES_Track_sigmas} where clearly the $\pi_{xy}\hat{X}$ closely accurately tracks $\pi_{xy}X(t)$ by adaptive tuning of $\mathbf{c}(t)$, resulting in accurate tracking of the entire 4D phase space which is quantified by accurate predictions of $\sigma_x$, $\sigma_{x'}$, $\sigma_y$, and $\sigma_{y'}$ throughout the drift process.

Figure \ref{fig:charge_state} shows accurate tracking of the time-varying charge state based only on matching $\pi_{xy}(t)$.

%% file: sec/5_conclusion.tex
\section{Conclusion}
\label{sec:conclusion}

We presented PhaseFlow4D, a feedback-guided latent diffusion framework for physically constrained 4D phase space reconstruction and adaptive online tracking of charged particle beams.  By embedding projection consistency as a hard architectural prior in the 4D VAE decoder, and by coupling the conditional latent diffusion model to a model-free extremum-seeking feedback loop, PhaseFlow4D achieves accurate 4D phase space recovery from a single observable 2D projection stream — without retraining, without gradients, and without any model of the time-varying dynamics.  On multi-particle heavy-ion beam simulations at FRIB, PhaseFlow4D reproduces full 4D phase space distributions $11000\times$ faster than the reference physics simulator while faithfully tracking distribution shifts driven by time-varying ECR source conditions.

Our results demonstrate that the core principles developed in the visual generative reconstruction community — learned latent spaces, projection-consistent architectural priors, and adaptive conditioning at inference time — transfer robustly to scientific and engineering inverse problems where ``observations'' are physical projections rather
than camera images.  We anticipate that this approach generalizes beyond particle accelerators to any high-dimensional dynamical system that is only partially observable through lower-dimensional projections, including plasma diagnostics, medical tomography,
and fluid state estimation.

%% file: sec/ES_suppl.tex
\clearpage
\section{Appendix}
\label{sec:appendix}

\begin{figure*}[h]
  \centering
  \includegraphics[width=0.95\linewidth]{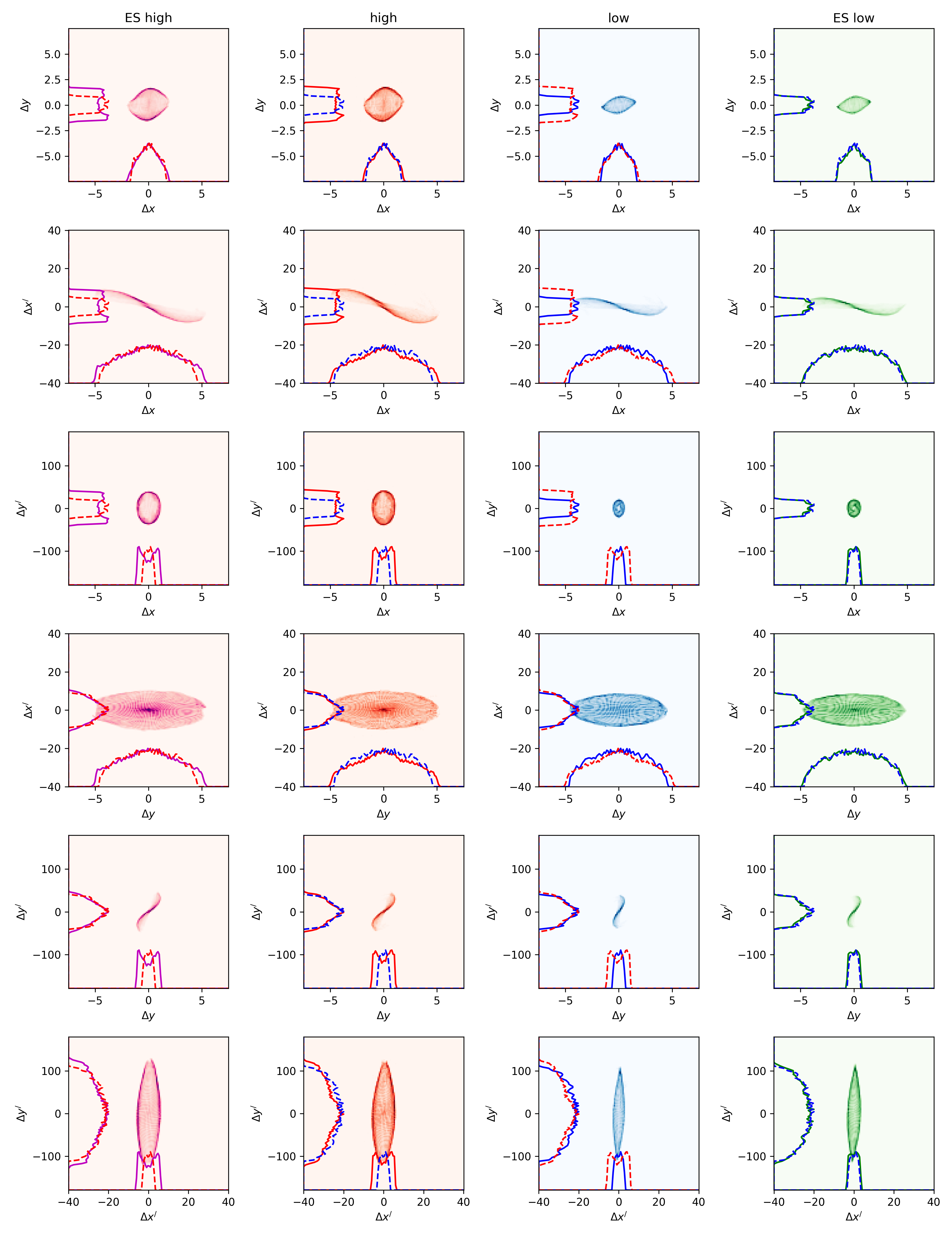}
  \caption{True and tracked projections of the beam are shown for various charge states during the ES-based tracking procedure.}
  \label{fig:ES_start_mid_end}
\end{figure*}

All six projections of the 4D phase space are shown for the lowest and highest level of charge neutralization at the start of the extremum seeking time-varying charge neutralization tracking problem in Figure \ref{fig:ES_start_mid_end}.

\begin{figure*}[h]
  \centering
  \includegraphics[width=0.95\linewidth]{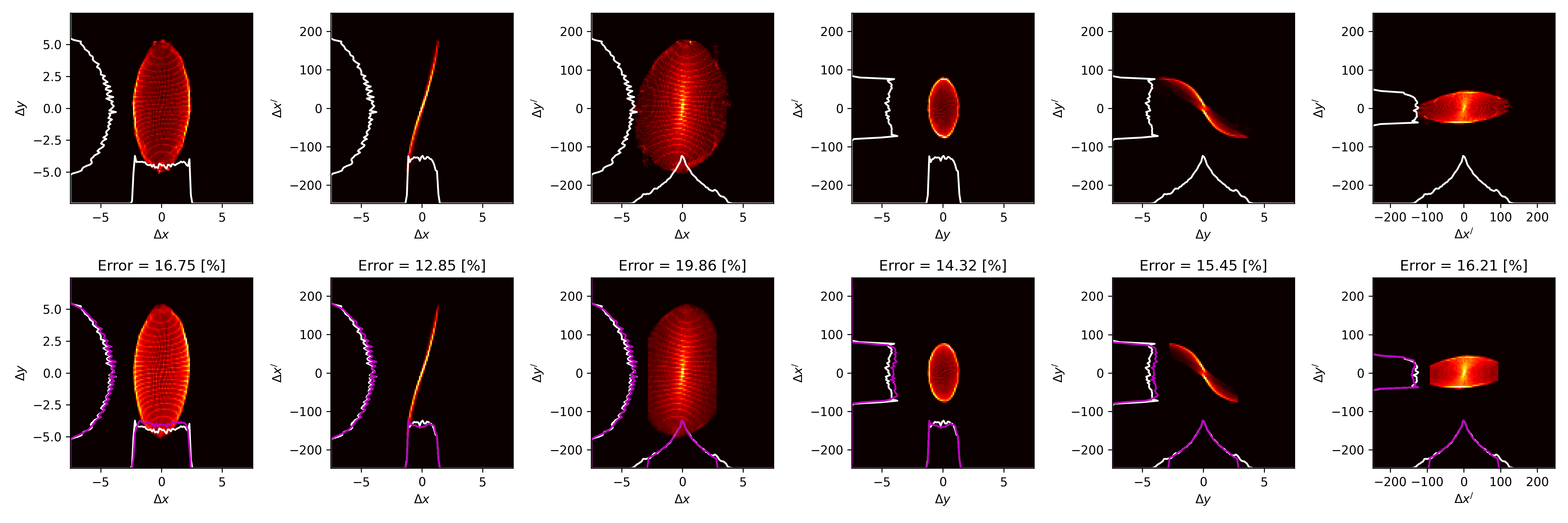}
  \caption{True and generated examples from test set.}
  \label{fig:ES_test1}
\end{figure*}

\begin{figure*}[h]
  \centering
  \includegraphics[width=0.95\linewidth]{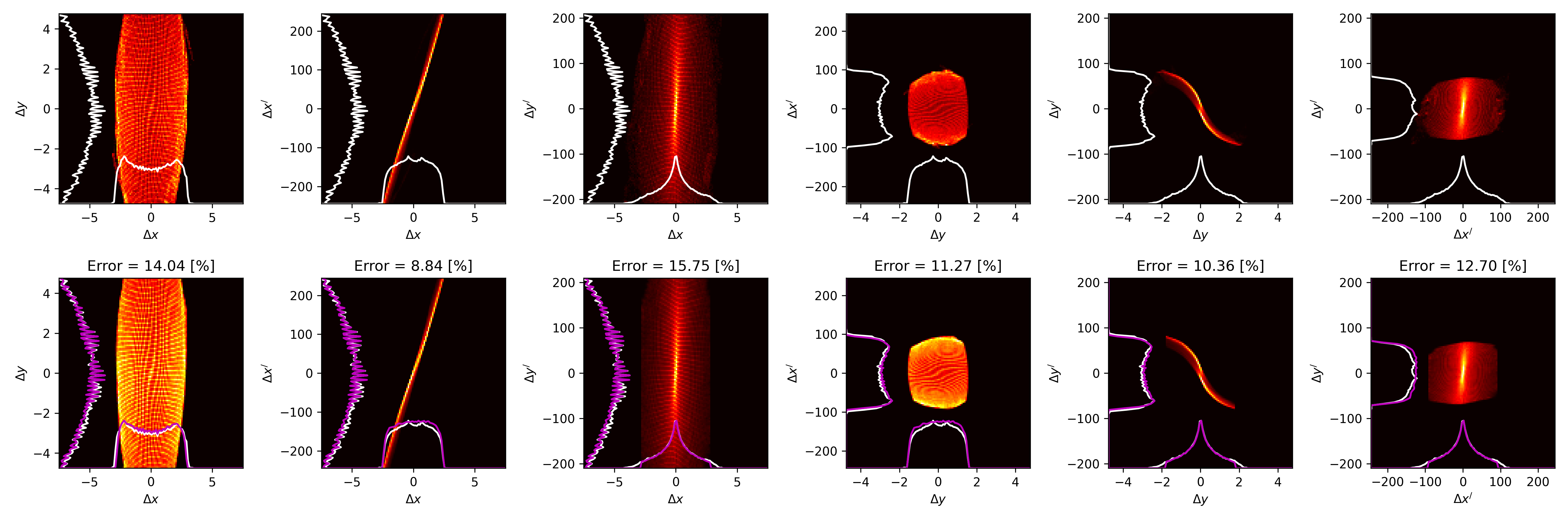}
  \caption{True and generated examples from test set.}
  \label{fig:ES_test2}
\end{figure*}

\begin{figure*}[h]
  \centering
  \includegraphics[width=0.95\linewidth]{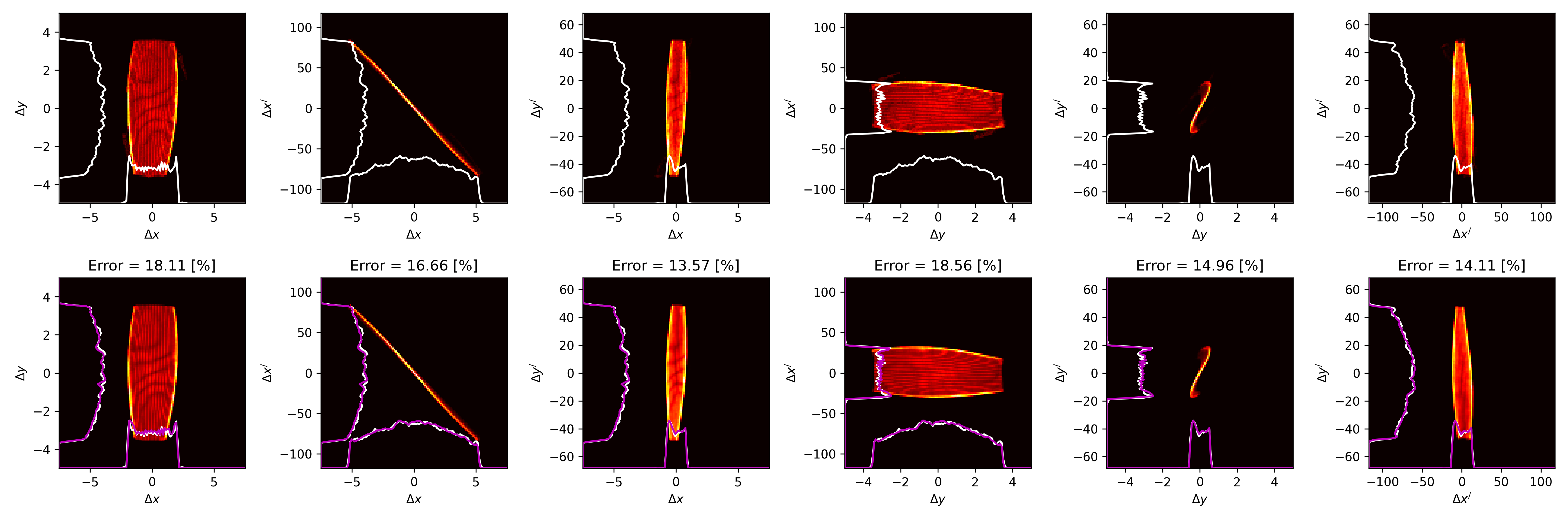}
  \caption{True and generated examples from test set.}
  \label{fig:ES_test3}
\end{figure*}

\begin{figure*}[h]
  \centering
  \includegraphics[width=0.95\linewidth]{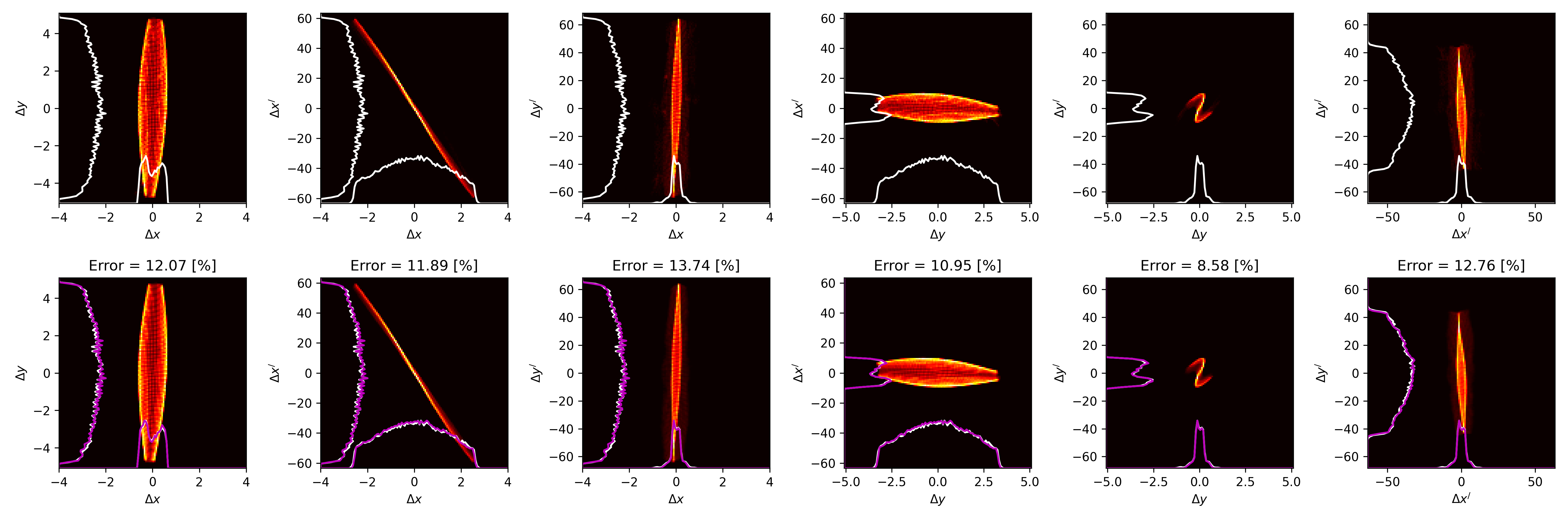}
  \caption{True and generated examples from test set.}
  \label{fig:ES_test4}
\end{figure*}

\begin{figure*}[h]
  \centering
  \includegraphics[width=0.95\linewidth]{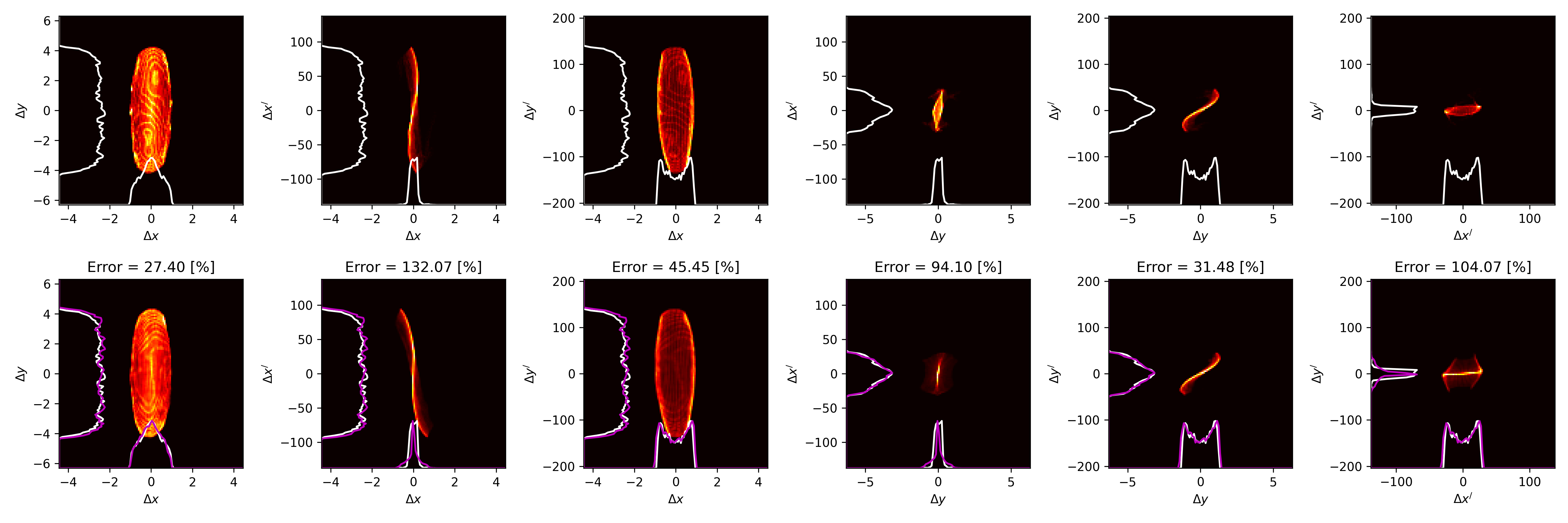}
  \caption{True and generated examples from test set.}
  \label{fig:ES_test5}
\end{figure*}

\begin{figure*}[h]
  \centering
  \includegraphics[width=0.95\linewidth]{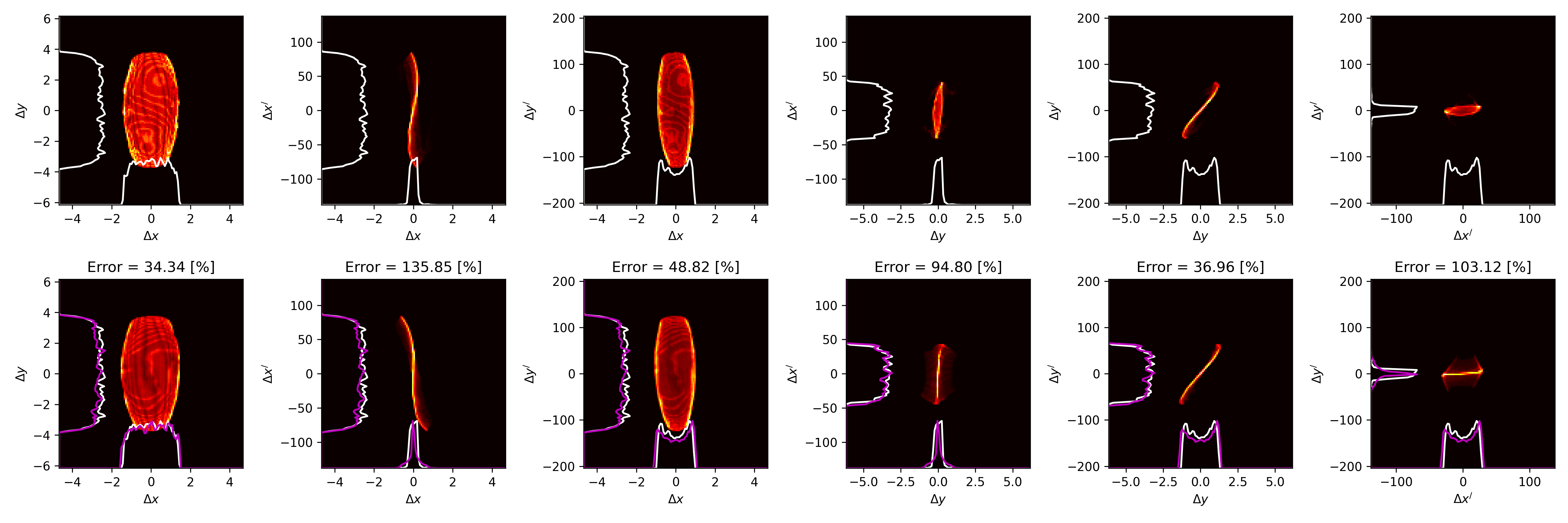}
  \caption{True and generated examples from test set.}
  \label{fig:ES_test6}
\end{figure*}

\begin{figure*}[h]
  \centering
  \includegraphics[width=0.95\linewidth]{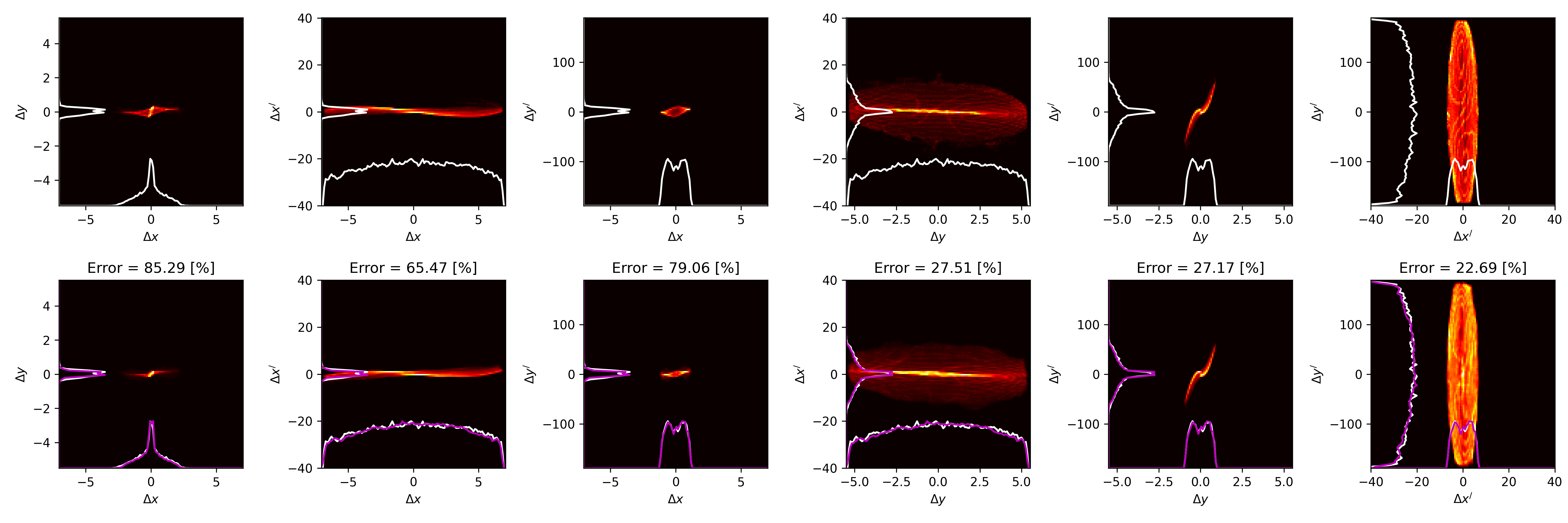}
  \caption{True and generated examples from test set.}
  \label{fig:ES_test7}
\end{figure*}

\begin{figure*}[h]
  \centering
  \includegraphics[width=0.95\linewidth]{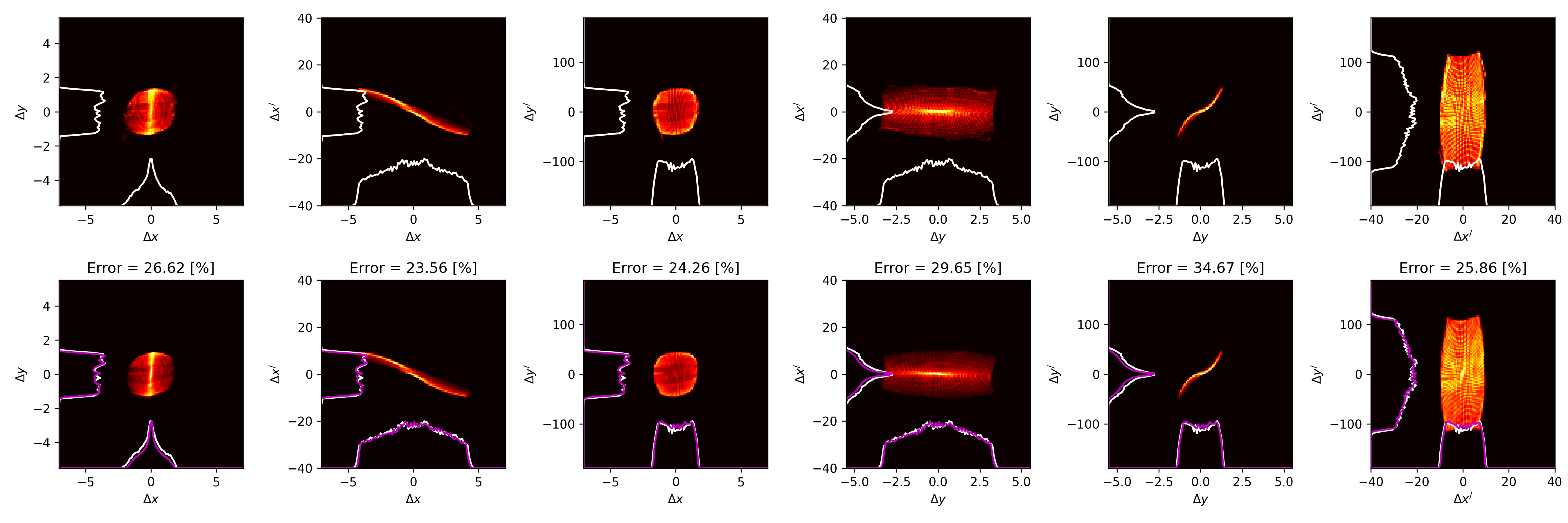}
  \caption{True and generated examples from test set.}
  \label{fig:ES_test8}
\end{figure*}

\begin{figure*}[h]
  \centering
  \includegraphics[width=0.95\linewidth]{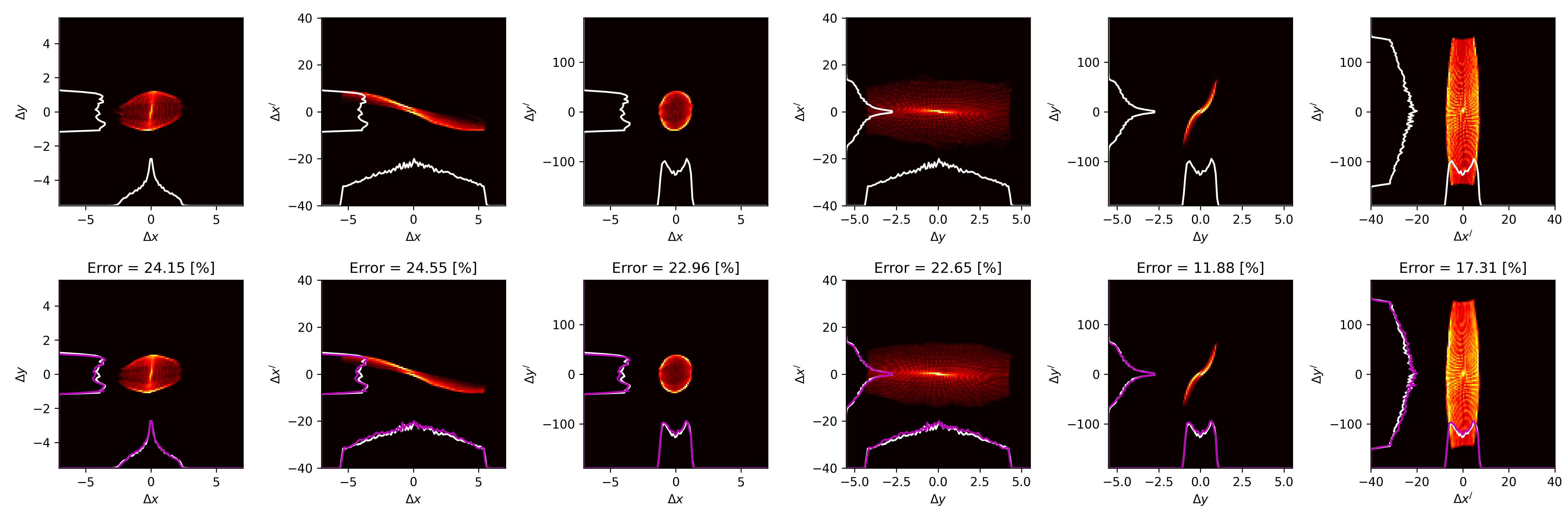}
  \caption{True and generated examples from test set.}
  \label{fig:ES_test9}
\end{figure*}

\begin{figure*}[h]
  \centering
  \includegraphics[width=0.95\linewidth]{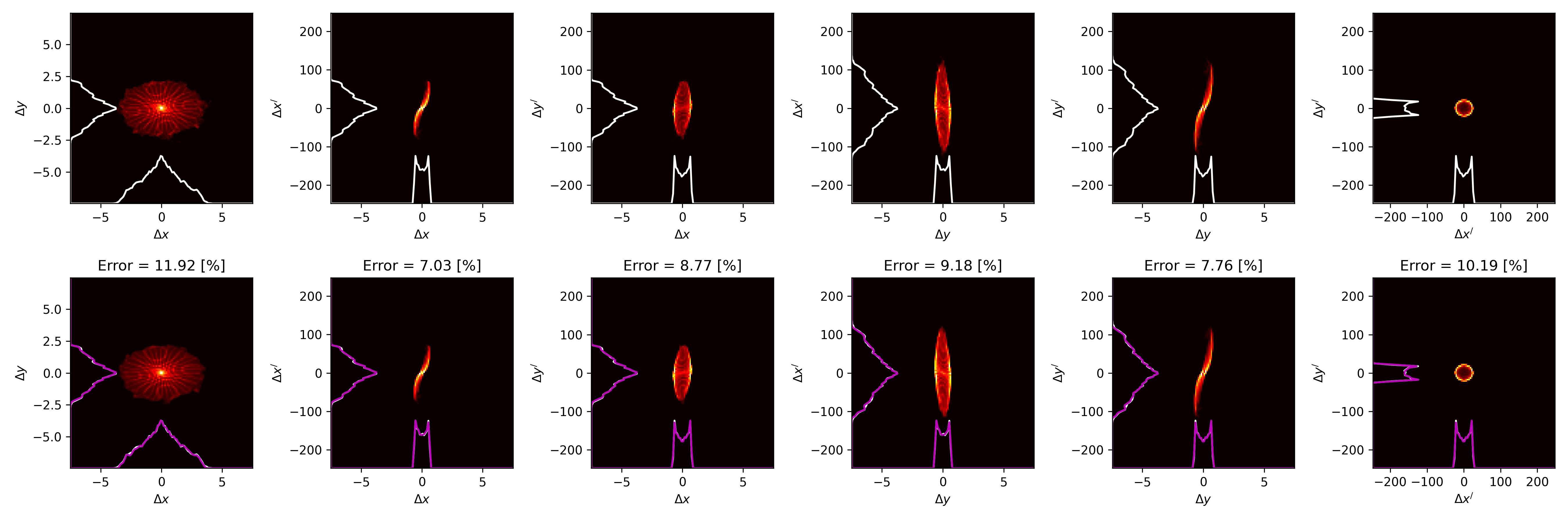}
  \caption{True and generated examples from test set.}
  \label{fig:ES_test10}
\end{figure*}

Figures \ref{fig:ES_test1}-\ref{fig:ES_test10} show detailed comparisons of random test data reconstructions.

%% file: main.bib
@String(CVPR= {IEEE Conf. Comput. Vis. Pattern Recog.})

@String(ECCV= {Eur. Conf. Comput. Vis.})

@String(CVPR  = {CVPR})

@String(ECCV  = {ECCV})

@article{scheinker2024cdvae,
  title={c{DVAE}: {VAE}-guided diffusion for particle accelerator beam 6{D} phase space projection diagnostics},
  author={Scheinker, Alexander},
  journal={Scientific Reports},
  volume={14},
  number={1},
  pages={29303},
  year={2024},
  publisher={Nature Publishing Group UK London}
}

@inproceedings{ref_ES_1,
  title={Model independent beam tuning},
  author={Scheinker, Alexander},
  booktitle={Int. Particle Accelerator Conf.(IPAC'13), Shanghai, China, 19-24 May 2013},
  pages={1862--1864},
  year={2013},
  organization={JACOW Publishing, Geneva, Switzerland},
  url={http://accelconf.web.cern.ch/AccelConf/IPAC2013/papers/tupwa068.pdf?n=IPAC2013/papers/tupwa068.pdf}
}

@article{scheinker2025physics,
  title={Physics-constrained superresolution diffusion for six-dimensional phase space diagnostics},
  author={Scheinker, Alexander},
  journal={Physical Review Research},
  volume={7},
  number={2},
  pages={023091},
  year={2025},
  publisher={APS}
}

@article{kerbl20233d,
  title={3{D} {G}aussian splatting for real-time radiance field rendering.},
  author={Kerbl, Bernhard and Kopanas, Georgios and Leimk{\"u}hler, Thomas and Drettakis, George and others},
  journal={ACM Trans. Graph.},
  volume={42},
  number={4},
  pages={139--1},
  year={2023}
}

@inproceedings{chen1993view,
title           ={View Interpolation for Image Synthesis},
author          ={Chen, Shenchang Eric and Williams, Lance},
booktitle       ={Proceedings of the 20th Annual Conference on Computer Graphics and Interactive Techniques (SIGGRAPH'93)},
pages           ={279–288},
year            ={1993},
organization    ={ACM}
}

@inproceedings{levoy1996light,
title           ={Light Field Rendering},
author          ={Levoy, Marc and Hanrahan, Pat},
booktitle       ={Proceedings of the 23rd Annual Conference on Computer Graphics and Interactive Techniques (SIGGRAPH'96)},
pages           ={31–42},
year            ={1996},
organization    ={ACM}
}

@inproceedings{gortler1996lumigraph,
title           ={The lumigraph},
author          ={Gortler, Steven J. and Grzeszczuk, Radek and Szeliski, Richard and Cohen, Michael F.},
booktitle       ={Proceedings of the 23rd Annual Conference on Computer Graphics and Interactive Techniques (SIGGRAPH'96)},
pages           ={43–54},
year            ={1996},
organization    ={ACM}
}

@article{nerf2021,
  title     ={Nerf: Representing scenes as neural radiance fields for view synthesis},
  author    ={Mildenhall, Ben and Srinivasan, Pratul P and Tancik, Matthew and Barron, Jonathan T and Ramamoorthi, Ravi and Ng, Ren},
  journal   ={Communications of the ACM},
  volume    ={65},
  number    ={1},
  pages     ={99--106},
  year      ={2021},
  publisher ={ACM New York, NY, USA}
}

@inproceedings{rombach2022ldm,
  title     = {High-Resolution Image Synthesis with Latent Diffusion Models},
  author    = {Rombach, Robin and Blattmann, Andreas and Lorenz, Dominik and Esser, Patrick and Ommer, Bj{\"o}rn},
  booktitle = {IEEE/CVF Conference on Computer Vision and Pattern Recognition (CVPR)},
  pages     = {10684--10695},
  year      = {2022}
}

@inproceedings{ho2022cfg,
  title     = {Classifier-Free Diffusion Guidance},
  author    = {Ho, Jonathan and Salimans, Tim},
  booktitle = {NeurIPS 2021 Workshop on Deep Generative Models and Downstream Applications},
  year      = {2021}
}

@inproceedings{sparf2023,
  title     = {{SPARF}: Neural Radiance Fields from Sparse and Noisy Poses},
  author    = {Truong, Prune and Rakotosaona, Marie-Julie and Manhardt, Fabian and Tombari, Federico},
  booktitle = {IEEE/CVF Conference on Computer Vision and Pattern Recognition (CVPR)},
  pages     = {4190--4200},
  year      = {2023}
}

@inproceedings{p2pbridge2024,
  title     = {{P2P-Bridge}: Diffusion Bridges for {3D} Point Cloud Denoising},
  author    = {Vogel, Mathias and Tateno, Keisuke and Pollefeys, Marc and Tombari, Federico and Rakotosaona, Marie-Julie and Engelmann, Francis},
  booktitle = {European Conference on Computer Vision (ECCV)},
  year      = {2024}
}

@inproceedings{flowr2025,
  title     = {{FlowR}: Flowing from Sparse to Dense {3D} Reconstructions},
  author    = {Fischer, Tobias and Bul{\`o}, Samuel Rota and Yang, Yung-Hsu and Keetha, Nikhil and Porzi, Lorenzo and M{\"u}ller, Nikolas and Schwarz, Katja and Luiten, Jonathon and Pollefeys, Marc and Kontschieder, Peter},
  booktitle = {IEEE/CVF Conference on Computer Vision and Pattern Recognition (CVPR)},
  year      = {2025}
}

@inproceedings{difix2025,
  title={Difix3d+: Improving 3d reconstructions with single-step diffusion models},
  author={Wu, Jay Zhangjie and Zhang, Yuxuan and Turki, Haithem and Ren, Xuanchi and Gao, Jun and Shou, Mike Zheng and Fidler, Sanja and Gojcic, Zan and Ling, Huan},
  booktitle={Proceedings of the IEEE/CVF Conference on Computer Vision and Pattern Recognition},
  pages={26024--26035},
  year={2025}
}

@inproceedings{gen3c2025,
  title     = {{Gen3C}: {3D}-Informed World-Consistent Video Generation with Precise Camera Control},
  author    = {Ren, Xuanchi and Shen, Tianchang and Huang, Jiahui and Ling, Huan and Lu, Yifan and Nimier-David, Merlin and M{\"u}ller, Thomas and Keller, Alexander and Fidler, Sanja and Gao, Jun},
  booktitle = {IEEE/CVF Conference on Computer Vision and Pattern Recognition (CVPR)},
  year      = {2025}
}

@inproceedings{objectx2025,
  title     = {{Object-X}: Learning to Reconstruct Multi-Modal {3D} Object Representations},
  author    = {Di Lorenzo, Gaia and Tombari, Federico and Pollefeys, Marc and Barath, Daniel},
  booktitle = {Advances in Neural Information Processing Systems (NeurIPS)},
  year      = {2025}
}

@article{scheinker2017es,
   title   = {Bounded extremum seeking with discontinuous dithers},
   author  = {Scheinker, Alexander and Scheinker, David},
   journal = {Automatica},
   volume  = {69},
   pages   = {250--257},
   year    = {2016},
   publisher = {Elsevier}
}

@inproceedings{ma2025inference,
   title     = {Inference-Time Scaling for Diffusion Models beyond Scaling Denoising Steps},
   author    = {Ma, Nanye and Goldstein, Mark and Albergo, Michael S and Boffi, Nicholas M
                and Vanden-Eijnden, Eric and Xie, Saining},
   booktitle = {arXiv preprint arXiv:2501.09732},
   year      = {2025}
 }

@article{ho2020ddpm,
   title   = {Denoising diffusion probabilistic models},
   author  = {Ho, Jonathan and Jain, Ajay and Abbeel, Pieter},
   journal = {Advances in Neural Information Processing Systems},
   volume  = {33},
   pages   = {6840--6851},
   year    = {2020}
}

@misc{track_ref,
   title  = {{TRACK}: A Code for Beam Dynamics Simulations},
   author = {Ostroumov, P.~N. and others},
   note   = {Facility for Rare Isotope Beams, Michigan State University},
   year   = {2020}
}
